\documentclass[runningheads,a4paper]{llncs}

\usepackage{enumitem}
\setlist{nolistsep}

\usepackage{amssymb}
\usepackage{graphicx}
\usepackage[usenames,dvipsnames]{color} 

\usepackage{url}

\usepackage{amsmath}
\usepackage{textcomp} 
\usepackage{xspace} 
\usepackage{bbding} 
\usepackage{pifont} 
\usepackage{stmaryrd} 
\usepackage{mdwlist} 
\usepackage[cp1252]{inputenc}

\begin{document}
\newcommand{\epar}[1]{``#1''}

\title{ ``Boring formal methods" or\\  ``Sherlock Holmes deduction methods"?
} 
\author{Maria Spichkova}
\institute{RMIT University,  Australia \\  \email{maria.spichkova@rmit.edu.au} }
\maketitle

\begin{abstract} 
This paper provides an overview of common challenges in teaching of logic and formal methods to Computer Science and IT students. 
We discuss our experiences from the course \emph{IN3050: Applied Logic in Engineering},  
 introduced as a ``logic for everybody" elective course at at TU Munich, Germany, to engage pupils studying Computer Science, IT and engineering subjects on Bachelor and Master levels.  
Our goal was  to overcome the bias that logic and formal methods are not only very complicated but also  very boring to study and to apply. 
In this paper, we present the core structure of the course, provide examples of exercises and evaluate the course based on the students' surveys. \footnote{%
Preprint. Accepted to the Software Technologies: Applications and Foundations (STAF 2016). Final version published by Springer International Publishing AG.
}
\end{abstract}

\section{Introduction}

 Logic not only helps to solve complicated and safety-critical problems, but also disciplines the mind and helps to develop abstract thinking, which is very important for any area of Computer Science and Engineering. 
 Problems in teaching and learning the basic principles of logic lead to the lack of analytical skills and abstract thinking as well as to 
 the problems in understanding of Formal Methods (FMs).  
The disputes on teaching logic and FMs have been going on for a long time, 
but most lecturers teaching these subjects agree that  they face many challenges specific  to these subjects.  
Students are strongly focused on the direct  relevance of what they study to their daily practice, and are not interested to study more fundamental subjects, especially logic \cite{tavolato2012integrating,jeannette2000wing}. The main obstacle in this case is that the students cannot match logic and FMs (in contrary to Games Development, Programming, Testing, etc.) to real world problems. 
As  curricula becomes more practice-oriented, the mathematical background of the students becomes weaker which provides an additional obstacle in understanding of logic and FMs, cf. \cite{bjorner201440,crocker2003teaching,Zamansky2015HOFMteaching}.  
Also, many students have negative perceptions and even fear of courses that require dealing with complex mathematical notations. This is strongly related to the phenomenon of \emph{mathematical anxiety}  \cite{MathsAnx_Wang,sherman2003mathematics}.  
The term  \emph{mathematical anxiety} was introduced in 1972 by Richardson and Suinn as
\emph{``feelings of tension and anxiety that interfere with the manipulation of numbers
and the solving of mathematical problems in a wide variety of ordinary life and
academic situations,"}~\cite{richardson1972mathematics}.
As stressed by Wang et al.,  mathematical anxiety has attracted
recent attention because of its damaging psychological effects and potential associations with mathematical problem
solving and achievement. 
From our point of view, this term could be extended to \emph{mathematical and logical anxiety} (or even to \emph{formal methods anxiety}), to cover a similar phenomenon on learning logic and FMs.

The term ``formal" is for many people just some kind of synonym for ``unreadable", however, even small syntactical changes of a formal method can make it more understandable and usable for an average engineer. 
In the course \emph{IN3050: Applied Logic in Engineering} we aimed to apply the core principles of our research work on \emph{Human Factors of Formal Methods} \cite{hffm_spichkova,spichkova2013design}, applying  the engineering psychology achievements to the design of FMs.
However, improving the usability aspects we cannot overcome the preconceived notions about FMs completely. 
To achieve the goal, we should start  not only by presenting theoretical aspects but also focusing on real applications, industrial and non-industrial ones, referring to the programming languages where the formal side is almost covered, or to famous fiction books and movies, e.g., to the famous  crime stories by A.C. Doyle. 
We applied these ideas within   the course \emph{IN3050: Applied Logic in Engineering} for Bachelor and Master students, and the students' feedback on this matter was very positive.  
There also is a great diversity in the students' background and cognitive skills due to the globalisation of higher education, which requires constant adaptation, cf. \cite{TNEHoare,TNEcrises}. One possible solution to overcome this problem is to provide courses that require very basic or  even no background knowledge in the corresponding areas, having as result a ``course for everybody", and providing students with deeper background additional non-compulsory tasks.

\emph{Contributions:}
Our goal was to overcome these problems and to teach the course \emph{IN3050:  Applied Logic in Engineering} 
without expecting any previous knowledge on logics and abstract thinking (in contrary to the many courses on logic and FMs). 
We introduced this lecture course as a ``logic for everybody", to engage pupils studying Computer Science (CS), IT and engineering subjects, 
to overcome the bias that logic and formal methods are not only very complicated but also  \epar{very boring to study and to apply}. 
As per evaluation report \cite{ace201213}, the majority of the students agreed that the course was helpful to their understanding of application of logic and FMs in Engineering.
We believe that this course would be especially beneficial for CS students, as well as for the IT students who aim to work as Software Requirements Engineers and Software Testers. 
A general introduction to this course was presented   in a technical report \cite{ALEtechreport}. In this paper we are going to focus on generalisation and analysis of the proposed solutions to improve students' learning experience.  
 
\emph{Outline:} 
The rest of the paper is organised as follows. 
Section \ref{sec:related} presents a short overview of the related work on teaching logic and FMs. 
Section \ref{sec:ale} introduces the core structure of the course \emph{IN3050: Applied Logic in Engineering}, where Section \ref{sec:exercises} presents a number of examples we used at the lectures and tutorials. 
Section \ref{sec:conclusions} concludes the paper evaluating the course based on the results from the students'
 surveys.

\section{Related Work}
\label{sec:related}

A symposium to explore and discuss the challenges and successful solutions in teaching of FMs was organised in 2004.
After 12 years, the lecturers face very similar problems while teaching logic and FMs: 
mathematical and logical anxiety as well as understandability and readability of FMs. 
However, over the last few years there have been  number of interesting and promising approaches that we would like to discuss here.  
In our previous work \cite{TeachingFM4SE}, 
we discussed the common issues in teaching of FMs and logic, as well as reviewed   various approaches for teaching FMs for Software Engineering that have been proposed, and discuss how they address the above mentioned challenges. The focus of our analysis here is on 
the  {\em collaborative} and {\em communication} aspects of software development using formal methods and logical modelling. 

A novel way to attract students while teaching FMs was presented in  \cite{curzon2013teaching}.
 Within the engagement project \emph{cs4fn}, Computer Science for Fun,   
the authors taught logic and computing concepts using magic tricks, which inspired students to work with logical tasks. 
Our approach was less revolutionary: we based the course on both practical examples and entertainment examples, such as formal modelling of logical puzzles and the Sherlock Holmes deductions from the modern BBC TV series \epar{Sherlock}.
 
Noble et al.~\cite{Noble2008} presented a course on  \emph{Introduction to Software Modelling}, where Alloy programming language was taught along with introduces the principles and practices of Software Engineering, beginning with domain analysis, specification of classes and use cases, writing invariants, etc. 
An interesting point about this douse is that the Alloy tool itself and the Alloy language were not introduced until the final two blocks of the course,
to allow focusing on software modelling, rather than on the technical tools.
 
 Wang and  Yilmaz suggested to group the study programs in three main categories, based on the way logic and FMs are integrated into software engineering curriculum, cf. \cite{Wang_strategy}: 
programs avoiding FMs, 
programs having  a specific course with emphasis on formal verification of source code, and 
programs redesigned to have FMs   integrated throughout the curriculum. 
This grouping does not cover another category, which we see as a very promising for integrating logic FMs into software engineering curriculum: 
 to introduce a specific course that 
 \begin{itemize}
\item[(1)] covers basics of logic and FMs, without requiring a deep knowledge in mathematics,  and
\item[(2)] uses visualisation and gamification/puzzle strategies to make the material more understandable and less boring for the students.
\end{itemize}
Examples of this kind of courses might be  
\begin{itemize}
\item
the \emph{Logic and FM} course designed for Information Systems students \cite{zamansky2015teaching}, 
\item
a series of courses specifically adapted to the needs of university of applied sciences, described in \cite{tavolato2012integrating},
\item 
Courses \emph{Computational Thinking}  at the Singapore Management University%
and
\emph{Computational Thinking and Design}  at the University of Maryland, %
organised in the spirit of ``computational thinking for everybody" envisioned by Wing  \cite{wing2006computational}.  
\end{itemize}
 The course \emph{IN3050: Applied Logic in Engineering}, which we introduced as 
a ``logic for everybody" course, can be seen as another example of this kind of courses.

\section{Course: Applied logic in Engineering}
\label{sec:ale}

The course \emph{IN3050: Applied logic in Engineering} (ALE) was introduced at TU Munich, Germany, in Winter Semester 2012/2013 as a face-to-face  course on Bachelor and Master levels.\footnote{\url{http://www4.in.tum.de/lehre/vorlesungen/Logic/WS1213/index.shtml}} 
The course was designed as an elective without any enforced prerequisites. It contributed 6 credit points to the student curriculum, 
which corresponds to 4 teacher-directed hours. 

In the case of ALE, the teacher-directed hours were divided into weekly lectures (2h a semester week) 
and weekly tutorials (2h a semester week). 
The course attracted 20 students from the following study programs:
\begin{itemize}
\item Computer Science (German, ``Informatik"),
\item Business Informatics (German, ``Wirtschaftsinformatik"), 
\item Mechanical Engineering (German, ``Maschinenwesen"). 
\end{itemize}
~\\
Introductory courses on Modelling in/for Software Engineering are usually taught in the first or second semester of the  first year  year of study.  
In contrast to this kind of courses, we  
\begin{itemize}
\item
focused not on principles and practices of Software Engineering, but on logical concepts, representation and analysis of information and problems;
\item
provided the course without any restriction on the year of study, and as result most of the students enrolled into this course  were either at the beginning of their study (1-3 semester) or at their final semesters (7th semester or later).
\end{itemize}
~\\
The exam for this course was organised as an \emph{open book} exam, as our goal was to examine whether the students understand and are able to apply the core principles of logic methods, rather than check they memory. 

The \emph{learning outcomes} of this course are that on completion of this course students 
\begin{itemize}
\item[(1)] 
will be able to state the basic principles of logic applied in Engineering,  and 
\item[(2)] 
will experience practical applications of these principles.
\end{itemize}

\begin{figure}[ht!]
\begin{center}
\includegraphics[scale=0.6]{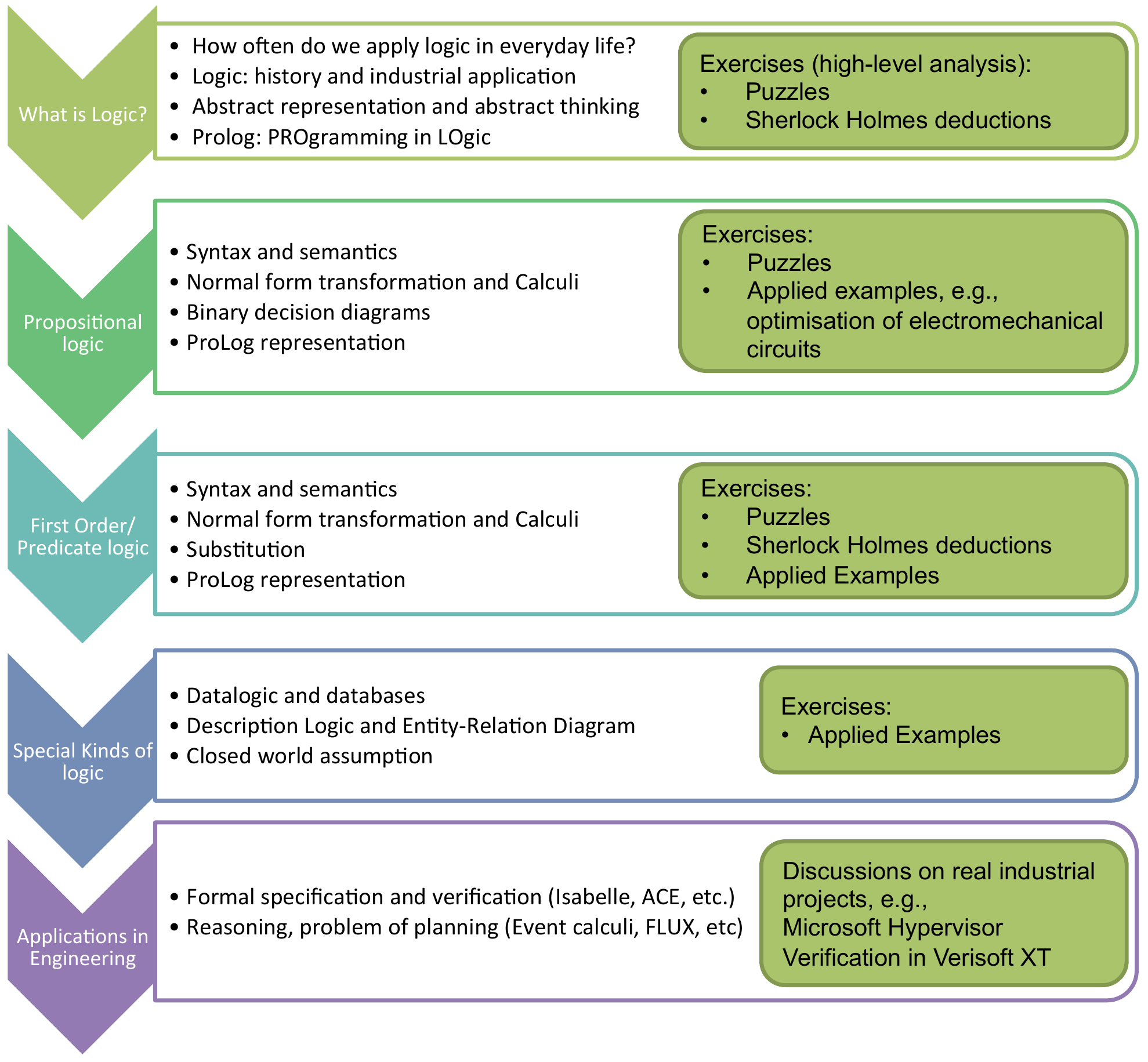}
\end{center}
\caption{Structure of the course \emph{Applied logic in Engineering} }
\label{fig:structure}
\end{figure}

~\\
The general structure of the course is presented on Figure \ref{fig:structure}. 
ALE is partially based on the book of Sch{\"{o}}ning \cite{Logic4CS},  which introduces the notions and methods of formal logic from a computer science standpoint, as well as on the book of  Russell and Norvig \cite{AIbook}.
 We also recommended our students to read the textbook of  Harrison~\cite{practLogic}, which focuses on practical application of logic and automated reasoning \cite{practLogic}, as well as a number of other books on logic and (semi-)automated theorem proving \cite{LogicCS,FOL_ATP,aussagenlogik}.
 
\newpage
\noindent
To explain the core ideas of Propositional Logic, First Order Logic (FOL) as well as of the special kinds of logics 
(such as  Datalogic, Description logic, etc.), we provided illustrative examples and exercises that were based both 
\begin{itemize}
\item
on application of the logics in Engineering, coming from real industrial problems,
\item
on puzzles and analysis of situations from famous fiction books and movies, e.g., 
detective stories like the famous Sherlock Holmes  crime stories written by A.C.~Doyle. 
\end{itemize}
The second kind of examples and exercises was required to provide  
more entertainment background for the course and to illustrate that logic is not necessary  a very dry subject.

~\\
Thus, the course introduces not only the basic principles of Propositional and First Order logic, 
but also presents the applied nature of logic and FMs, such as 
\begin{itemize}
\item Reasoning and Planning problems;
\item Formal Specifications/ models for precise description of systems and requirements
and analysis of systems;
\item Verification: Proving that a system fulfils its requirements, and that 
 a new version of a system is a refinement of the previous version;
\item Theorem proving/Model checking allowing 
  (semi-)automated proofs;
\item  Design/optimization of digital circuits:
Claude Shannon  has shown that propositional logic can be used to
describe and optimize electromechanical circuits, \cite{Shannon1937};
 \item  Formalisation of queries in databases.
\end{itemize}
We also analysed application of FMs in a number of recent research projects, as well as discussed our experience from large scale industrial projects  involving FMs, focusing not only on the efficiency features but also on usability aspects and corresponding feedback from industrial partners \cite{spichkova2012verified,spichkova2006flexray,kuhnel2006upcoming,kuhnel2006flexray,feilkas2011refined,holzl2010autofocus,botaschanjan2008correctness,feilkas2009top}.

\section{Examples and exercises provided within the course}
\label{sec:exercises}

 In this section we discuss examples and exercises   introduced within the course. 
 
 \begin{figure}[ht!]
\begin{center}
\includegraphics[scale=0.45]{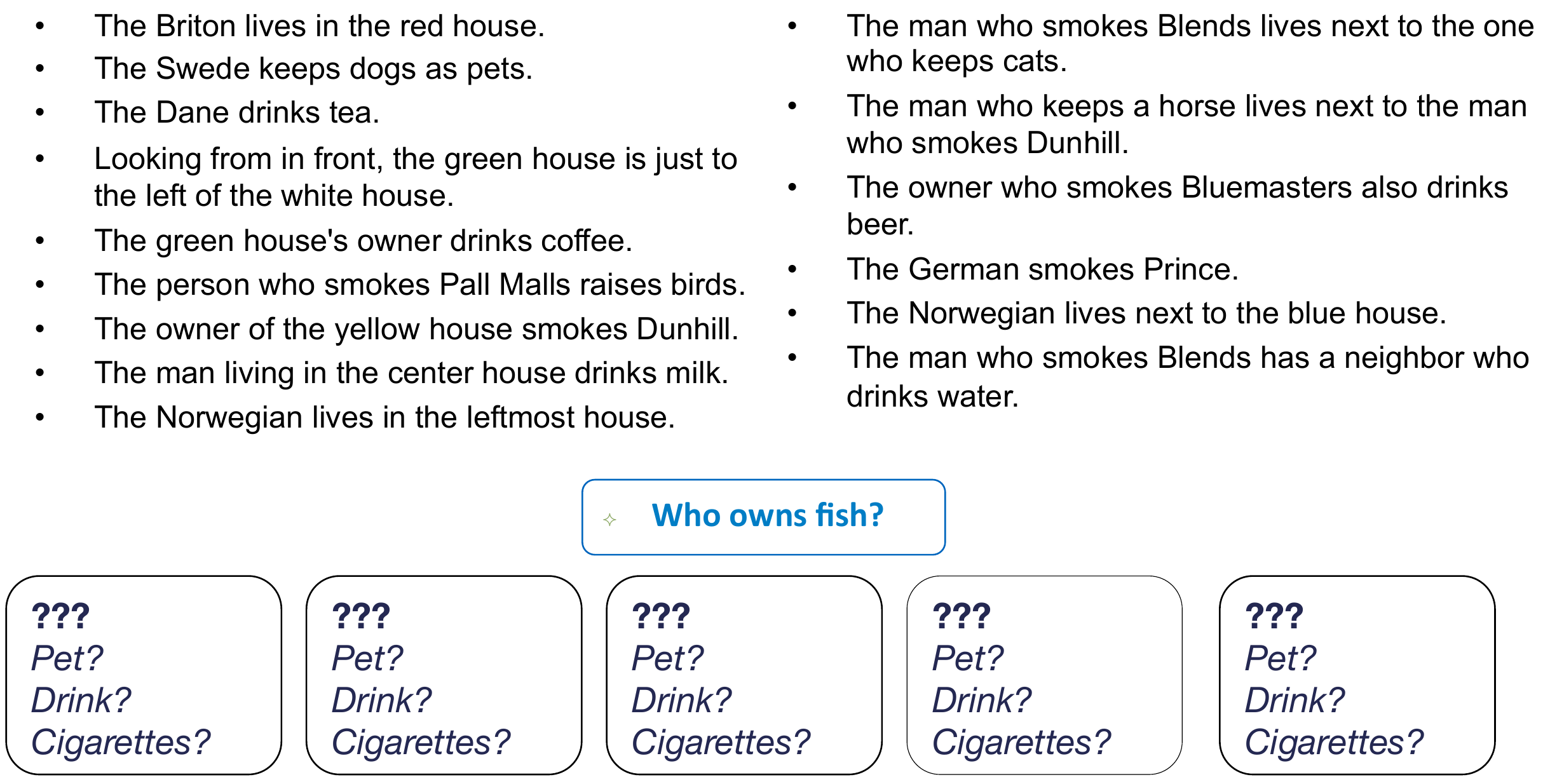}
\end{center}
\vspace{-5mm}
\caption{Solving the Einstein puzzle: Step 1}
\label{fig:einstein1}
\end{figure}

\begin{figure}[ht!]
\begin{center}
\includegraphics[scale=0.45]{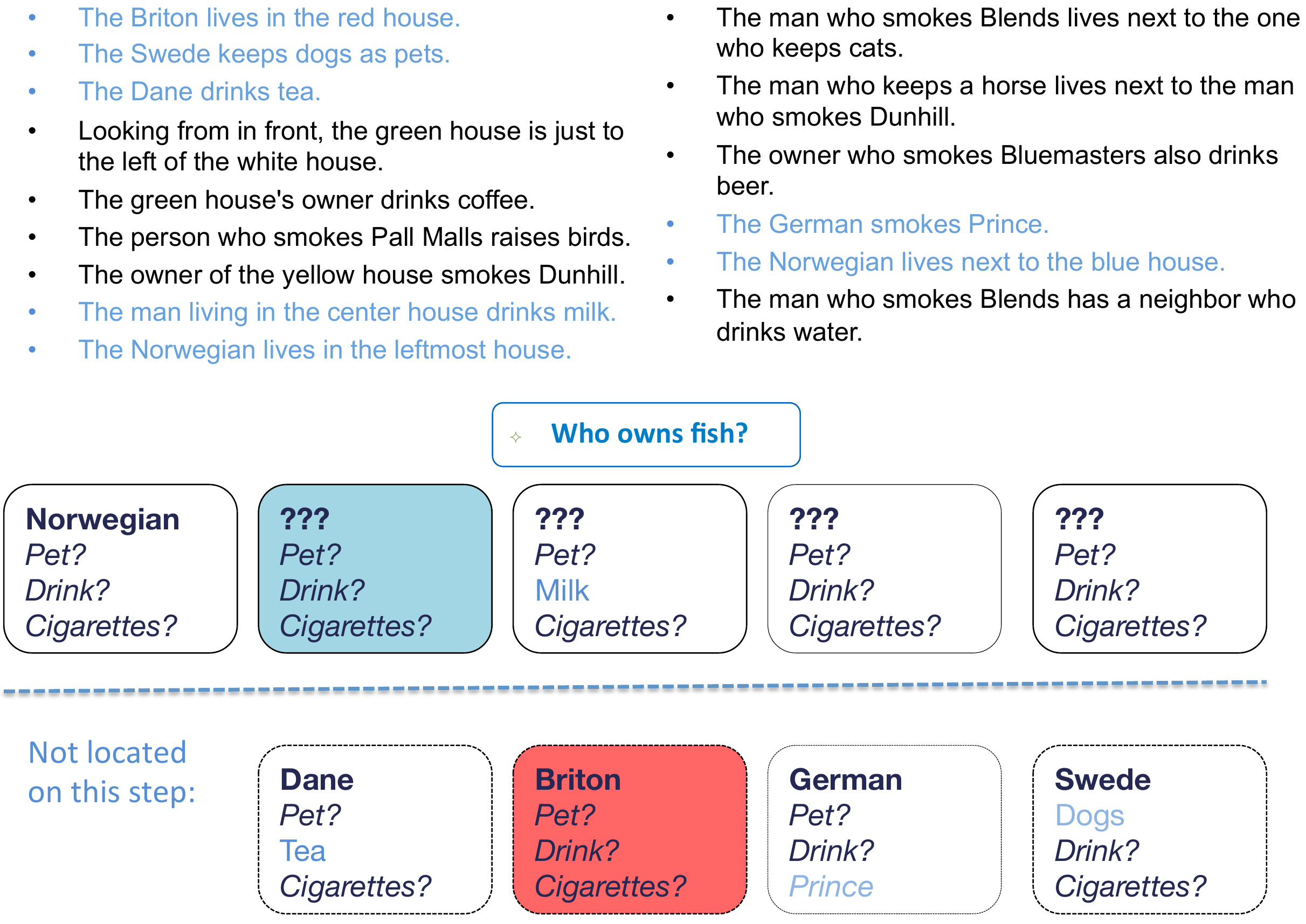}
\end{center}
\vspace{-5mm}
\caption{Solving the Einstein puzzle: Step 2}
\label{fig:einstein2}
\end{figure}

\begin{figure}[ht!]
\begin{center}
\includegraphics[scale=0.45]{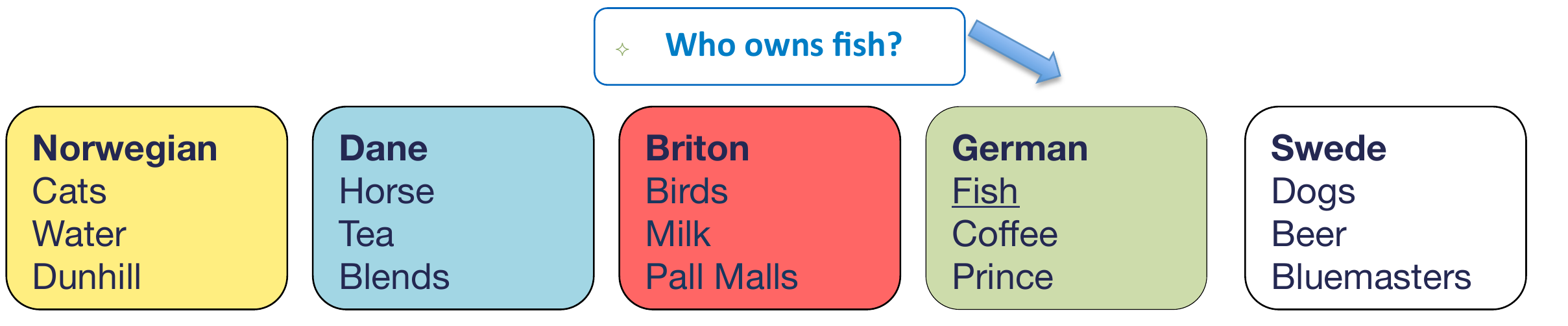}
\end{center}
\vspace{-5mm}
\caption{Solving the Einstein puzzle: Step 3}
\label{fig:einstein3}
\end{figure}

~\\ 
\textbf{Example:  Propositional Logic.} 
This example we used to explain visually how to solve a suggested by Einstein logical puzzle, also in 
Propositional Logic. 
Figure \ref{fig:einstein1} presents the  task of the puzzle and the initial set up for the suggested visual framework, where the five blocks represent the houses. 
In the second step, presented on Figure \ref{fig:einstein2}, we apply all the facts highlight hem with light blue, and visualise the corresponding information. 
In the next step we generate additional rules based on the facts we already know and solve the puzzle, as shown on Figure \ref{fig:einstein3}.

~\\
\textbf{Exercise: Applied Propositional Logic.}
\emph{
Formalise the following sentences $S_{1}$ and $S_{2}$ as formulas and then show that they are equivalent:
	\begin{itemize}
		\item[$S_{1}$:] 
		If the communication    fails or the battery power gets low,  while the system is in
 sending mode, then the system goes into safety mode.
		\item[$S_{2}$:]
		If the communication   
 fails, then the system must go into safety mode provided that
 it is in sending mode; and if it is in sending mode, it goes into safety mode, if the
 battery power gets low.
		\end{itemize}
		}
\noindent
To solve this task it is enough to apply Propositional Logic. We define the following four propositions to show that the above sentences are equivalent 

$A=$ \epar{communication fails} 

	$B=$ \epar{battery power gets low} 
	
	$C=$ \epar{system is in sending mode} 
	
	$D=$ \epar{system gets into safety mode}

\noindent	
Then we will have	
\[
\begin{array}{ll}
S_{1}: &   (A \vee B) \wedge C \longrightarrow D 
\\
 S_{2} & (A \longrightarrow (C \longrightarrow D)) \wedge (C \longrightarrow (B \longrightarrow D)) 
 \end{array}
\]
	First step: simplify $S_{1}$:
	\[
\begin{array}{l}
	 (A \vee B) \wedge C \longrightarrow D\equiv\\
	 \neg ((A \vee B) \wedge C) \vee D\equiv \\
	 \neg(A \vee B)\vee \neg C \vee D\equiv\\
	 (\neg A \wedge \neg B) \vee \neg C \vee D 
	 \end{array}
\]
 	Second step: simplify $S_{1}$:
		\[
\begin{array}{l}
 (A \longrightarrow (C \longrightarrow D)) \wedge (C \longrightarrow (B \longrightarrow D))\equiv  
 \\
 (\neg A \vee \neg C \vee D) \wedge (\neg C \vee \neg B \vee D)\equiv  
 \\
 \neg A \wedge (\neg C \vee \neg B \vee D) \vee \neg C \wedge (\neg C \vee \neg B \vee D) \vee D \wedge (\neg C \vee \neg B \vee D)\equiv
  \\
 (\neg A \wedge\neg C) \vee (\neg A \wedge \neg B) \vee (\neg A \wedge D) \vee (\neg C) \vee (\neg C \wedge \neg B) \vee (\neg C \wedge D)\\
 ~~~~~  \vee (D \wedge \neg C) \vee (D \wedge \neg B) \vee (D)\equiv
 \\
(\neg A \wedge\neg C)  \vee (\neg A \wedge \neg B) \vee   (\neg A \wedge D) \vee (\neg C) \vee  (\neg C \wedge \neg B)  \vee (\neg C \wedge D)\\
 ~~~~~    \vee  (D \wedge \neg C)  \vee  (D \wedge \neg B)  \vee (D)\equiv \\
 (\neg A \wedge \neg B) \vee \neg C \vee D 
	 \end{array}
\]
This proves semantical equivalence of the formulas.\\
 \hspace{111mm}$\Box$	

\begin{figure}[ht!]
\begin{center}
\includegraphics[scale=0.45]{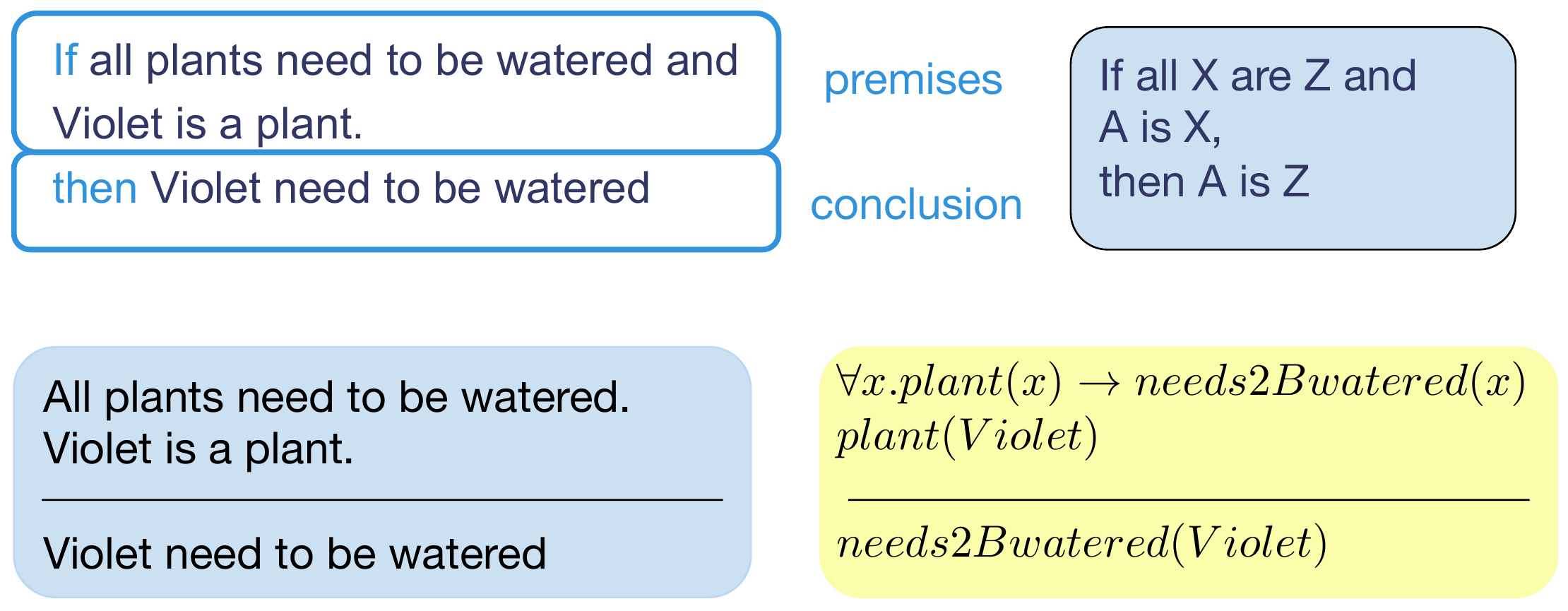}
\end{center}
\vspace{-5mm}
\caption{Visual explanation of formal notation: Introduction to the Syllogisms}
\label{fig:Syllogisms}
\end{figure}

 ~\\ ~\\ 
\noindent
\textbf{Example:  First Order Logic.} 
Figure \ref{fig:Syllogisms} provides an example we used to explain the idea of formal notation for syllogisms.\\
~

\newpage
\noindent
\textbf{Exercise: Applied First Order Logic.}
\emph{
	Formalize the following sentences as formulas and then show that 
	they are equivalent:
	\begin{itemize}
		\item[(1)] The following property holds not for all time intervals: 
		If the system gets a signal from its sensors that there is no communication  at a time interval $t$  or that  the 
					battery power gets low  at a time interval $t$, and exists an information package that have to be send, then at a time interval $t$
					there is an information package in the temporal buffer.
		\item[(2)] At some time interval $t$ the following holds for all information packages:
		there is an information package that have to be send, but
		there is no information package in the temporal buffer, and 
		the system gets a signal from its sensors that there is no communication   or that  the 
					battery power gets low.
	\end{itemize}
	}
\noindent
~\\
One possible solution:\\
Formalisation of the sentences would be	\\
(1) $\neg\forall t. \left( (C(t) \vee B(t)) \wedge S(t) \to T(t) \right)$  and\\
(2) $\exists t. \left(  S(t) \wedge \neg T(t) \wedge (C(t) \vee B(t))  \right)$.\\
Proof that both formulas are equal:
\\
$\neg\forall t. \left( (C(t) \vee B(t)) \wedge S(t) \to T(t) \right)$
\\
$\equiv \exists t. \neg \left( (C(t) \vee B(t)) \wedge S(t) \to T(t) \right)$
\\
$\equiv \exists t. \neg \left( \neg ( (C(t) \vee B(t)) \wedge S(t))  \vee T(t) \right)$
\\
$\equiv \exists t. \left(  ( (C(t) \vee B(t)) \wedge S(t))  \wedge \neg T(t) \right)$
\\
$\equiv \exists t. \left( S(t) \wedge \neg T(t) \wedge  (C(t) \vee B(t))  \right)$
\\[5mm]
Another possible solution:
\\
Formalization of (1): $\neg\forall t. \exists p. \left(  (C(t) \vee B(t)) \wedge S(p,t) \to T(p,t)  \right) $
\\
Formalization of (2): $\exists t. \forall p. \left(  S(p,t) \wedge \neg T(p,t) \wedge (C(t) \vee B(t))  \right)$
\\
Proof that both formulas are equal:
\\
$\exists t. \forall p. \left(  S(p,t) \wedge \neg T(p,t) \wedge (C(t) \vee B(t))  \right)$
\\
$\equiv \neg \forall t. \neg (\forall p. \left(  S(p,t) \wedge \neg T(p,t) \wedge (C(t) \vee B(t))  \right))$
\\
$\equiv \neg \forall t. (\exists p. \neg \left(  S(p,t) \wedge \neg T(p,t) \wedge (C(t) \vee B(t))  \right))$
\\
$\equiv \neg \forall t. (\exists p. \left(  \neg S(p,t) \vee T(p,t) \vee \neg (C(t) \vee B(t))  \right))$
\\
$\equiv \neg \forall t. (\exists p. \left(  \neg S(p,t) \vee \neg (C(t) \vee B(t)) \vee T(p,t)  \right))$
\\
$\equiv \neg \forall t. (\exists p. \left(  \neg (S(p,t) \wedge (C(t) \vee B(t))) \vee T(p,t)  \right))$
\\
$\equiv \neg \forall t. (\exists p. \left(  (S(p,t) \wedge (C(t) \vee B(t))) \to T(p,t)  \right))$

\hspace{111mm}$\Box$

\section{Evaluation and  Conclusions}
\label{sec:conclusions}

This paper presents an  overview of common challenges in teaching of formal methods and suggested solutions to them, based on
our experiences from the course \emph{Applied Logic in Engineering} taught at TU Munich, Germany.

The course was introduced as an elective course on Bachelor and Master levels and attracted 20 students. 
As per course evaluation~\cite{ace201213}, the majority of the students agreed that the provided examples were very helpful, and the learning amount and the amount of the material provided within the course  were ``exactly right" (German, ``genau richting"). 
For example, we received the following comments from our students:
\\
\emph{``Structured logically and builds up stuff part by part; nice additions as Sherlock video"};
\\
\emph{``The topic presented are interesting and indeed ``applied", unlike other logical courses that are more theoretic"};
\\
\emph{``I liked the small size of the course and I got a deeper understanding of logic"}.
To the question what did you most liked in the course, the students replied\\
\emph{``Sherlock, Examples during lecture"}. 

The students' feedback highlighted that the examples (for which we used visual representation to reduce the cognitive load of students and to introduce the corresponding ideas more understandable) as well as using 
 puzzles and situations from famous fiction books and movies, not only helps to understand the application of logic and FMs to real world problems, but also makes the leaning experience more interesting and helps to overcome the prejustice that the FMs are \emph{boring per default}. Another point that we took out from the evaluation report is that it would be beneficial for this kind of courses to have a  relatively small size of class, which allows teachers to approach each student individually.

\bibliographystyle{abbrv}

\end{document}